# THE MAGNITUDE OF THE EFFECT OF CALF MUSCLES FATIGUE ON POSTURAL CONTROL DURING BIPEDAL QUIET STANDING WITH VISION DEPENDS ON THE EYE-VISUAL TARGET DISTANCE

Nicolas VUILLERME [1,2], Cyril BURDET [1], Brice ISABLEU [3] and Sylvain DEMETZ [1]

[1] Laboratoire de Modélisation des Activités Sportives, Université de Savoie, France.

[2] Laboratoire TIMC-IMAG, UMR CNRS 5525, Equipe AFIRM, Grenoble, France.

[3] UFR STAPS – Centre de Recherche en Sciences du Sport (CRESS EA 1609), Université Paris-Sud XI, France.

Address for correspondence:

Nicolas VUILLERME

Laboratoire TIMC-IMAG, UMR UJF CNRS 5525

Faculté de Médecine

38706 La Tronche cédex

France.

Tel: (33) (0) 4 76 63 74 86

Fax: (33) (0) 4 76 51 86 67

Email: nicolas.vuillerme@imag.fr

# THE MAGNITUDE OF THE EFFECT OF CALF MUSCLES FATIGUE ON POSTURAL CONTROL DURING BIPEDAL QUIET STANDING WITH VISION DEPENDS ON THE EYE-VISUAL TARGET DISTANCE

**Abstract**

The purpose of the present experiment was to investigate whether, with vision, the magnitude of the effect of calf muscles fatigue on postural control during bipedal quiet standing depends on the eye–visual target distance. Twelve young university students were asked to stand upright as immobile as possible in three visual conditions (No vision, Vision 1 m and Vision 4 m) executed in two conditions of No fatigue and Fatigue of the calf muscles. Centre of foot pressure displacements were recorded using a force platform. Similar increased variances of the centre of foot pressure displacements were observed in the fatigue relative to the No fatigue condition for both the No vision and Vision 4 m conditions. Interestingly, in the vision 1 m condition, fatigue yielded: (1) a similar increased variance of the centre of foot pressure displacements to those observed in the No vision and Vision 4 m conditions along the medio-lateral axis and (2) a weaker destabilising effect relative to the No vision and Vision 4 m conditions along the antero-posterior axis. These results evidence that the ability to use visual information for postural control during bipedal quiet standing following calf muscles fatigue is dependent on the eye–visual target distance. More largely, in the context of the multisensory control of balance, the present findings suggest that the efficiency of the sensory reweighting of visual sensory cues as the neuro-muscular constraints acting on the subject change is critically linked with the quality of the information the visual system obtains.

1. Introduction

It is now well established that postural control of human beings is achieved by integrating sensory information from the visual, vestibular and somatosensory systems (e.g., [1]). Of these three inputs, the visual system is certainly the one that has received the most attention (e.g., [2], for a recent review). Our current understanding of visual stabilisation of postural sway is based on how effectively body oscillations can be detected relative to stationary environmental surfaces. A standard argument is that with increasing distance or decreasing size of a fixated target, the magnitude of the retinal displacement caused by the head sway during fixation of the target declines, yielding a deterioration of visual stabilisation (e.g., [3,4]). As a result, a nearby visual target has the potential to be more efficient in reducing postural sway than a target at some distance (e.g., [3–5]).

In recent years, the effects of localized calf muscles fatigue on postural control during bipedal quiet standing have received a growing interest [6–10]. While the abovementioned studies are conclusive in reporting a destabilising effect of fatigue in the absence of vision, this is not the case when visual information is available. On the one hand, Corbeil et al. [6] observed similar destabilizing effects of muscular fatigue with and without vision. On the other hand, Ledin et al. [7] showed that vision was able to substantially limit, along the antero-posterior axis mainly, the increased postural sway induced by the calf muscles fatigue.

At this point, it is important to mention that it seems rather difficult to make specific comparisons across experiments due to the nature of the fatigue protocol (bloc design-training program consisting of the execution of 100 repeated plantar-flexions starting at 75% of subjects' maximal workload with a reverse pyramidal technique in which the load was diminished gradually whenever subjects were unable to perform plantar-flexion + maximal isometric contractions [6] versus toe-lift exercise until maximal exhaustion repeated prior to each trial [7]), experimental procedure (modified Romberg protocol [6] versus normal stance with feet at an angle about 308 open to the front and the heals approximately 3 cm apart [7]), dependent variables analysed (e.g., centre of foot pressure (CoP) displacements [6] versus variance of body sway [7]) or the duration of the data analysed (60 s [6] versus 30 s [7]). It is possible these conflicting results to also stem from the different eye–visual target distances used in these experiments. Indeed, when vision was available, while ''subjects were instructed to fixate a point located 4m in front of them'' (p. 93) in the Corbeil et al. experiment [6], they ''focused on a mark placed on the wall about 1.5 m in front of the subject'' (p. 185) in Ledin et al. experiment [7].

Within this context, the purpose of the present experiment was to investigate whether, with vision, the magnitude of the effect of calf muscles fatigue on postural control during quiet standing depends on the distance to the visual target. Considering that (1) the influence of visual information on postural control decreases as the eye–visual target distance increases (e.g., [3–5]) and (2) the abovementioned results of Ledin et al. [7] and Corbeil et al. [6], we expected that (1) a nearby visual target would allow individuals to reduce postural sway following calf muscles fatigue with a stabilising effect mostly occurring in the sagittal plane [7], whereas a distant visual target would not [6].

## 2. Methods

### 2.1. Subjects

Twelve university students from the Department of Sports Sciences at the University of Savoie (age = 21.9 ± 1.7 years; body weight = 63.2 ± 6.6 kg; height = 173.1 ± 6.8 cm; mean ± S.D.) with no history of injury or pathology to either lower extremity and normal or corrected-to normal visual acuity voluntarily participated in the experiment. They gave their informed consent to the experimental procedure as required by the Helsinki declaration (1964) and the local Ethics Committee.

**2.2. Apparatus**

A force platform (Equi+, model PF01, Aix les Bains, France), constituted of an aluminium plate (800 mm each side) laying on three uniaxial load cells, was used to measure the CoP displacements. Signals from the force-platform were sampled at 64 Hz, amplified and converted from analogue to digital form.

**2.3. Task and procedure**

Subjects stood barefoot on the force platform in a natural position (feet abducted at 30°, heels separated by 3 cm), their arms hanging loosely by their sides. Subject's task was to sway as little as possible in three visual conditions of No vision, Vision 1 m and Vision 4 m. In the No vision condition, they were asked to close their eyes and to keep their gaze in a straight-ahead direction. In the Vision 1 m and Vision 4 m conditions, they were asked to stare at the intersection of a black cross (20 cm × 25 cm) placed onto the white wall distant 1 m and 4 m in front of them, at the eyes level, respectively.

These three visual conditions were performed under two experimental conditions. The No fatigue condition served as a control condition. For each visual condition (No vision, Vision 1 m and Vision 4 m), subjects performed three 32-s trials. The order of presentation of these three visual conditions was randomised over subjects. In the fatigue condition, the measurements were performed immediately after a fatiguing procedure. Its aim was to induce a muscular fatigue in the ankle plantar-flexor of both legs until maximal exhaustion, using a protocol similar to that used by [7]. Subjects were asked to perform toe-lifts as many times as possible following the beat of a metronome (40 beats/min). Verbal encouragement was given to ensure that subjects worked maximally. The fatigue level was reached when subjects were no more able to complete the exercise. To ensure that balance measurement in the Fatigue condition was obtained in a real fatigued state: (1) the fatiguing exercise took place beside the force platform to minimise the time between the exercise-induced fatiguing exercise and the measurement of the CoP displacements and (2) the fatiguing exercise was repeated prior to each trial [7–10]. Three additional trials for each visual condition were executed, for a total of 18 trials.

**2.4. Data analysis**

CoP displacements were processed through a space-time domain analysis including the calculation of the variances of positions along the medio-lateral and antero-posterior axes. This dependent variable provides a measure of amplitude variability of the CoP displacements around the mean position, over the sampled period, along each axis.

**2.5. Statistical analysis**

The means of the three trials performed in each of the six experimental conditions were used for statistical analyses. Two fatigues (No fatigue versus Fatigue) × three three visions (No vision versus Vision 1 m versus Vision 4 m) × two axes (Medio-lateral versus Antero-posterior) analyses of variance with repeated measures on all factors were applied to the data. Post hoc analyses (*L.S.D.* test) were used whenever necessary. Level of significance was set at 0.05.

**3. Results**

Analysis of the variance of the CoP displacements showed main effects of Fatigue ($F_{(1,11)} = 10.93$, $P < 0.01$), Vision ($F_{(1,22)} = 5.83$, $P < 0.01$) and Axis ($F_{(1,11)} = 5.34$, $P <$ 0.05). It also showed a significant two way-interaction of Vision × Axis (F (1,22) = 3.47, P < 0.05) and a significant three-way interaction of Fatigue × Vision × Axis (F (1,22) = 3.49, P < 0.05).

As illustrated in Fig. 1, the decomposition of the three-way interaction into its simple main effects indicated similar increased variances of the CoP displacements in the Fatigue relative to the No fatigue condition for both the No vision (left panel) and Vision 4 m conditions (right panel) (Ps < 0.01 and <0.001, for the medio-lateral and anteroposterior axes, respectively). In the Vision 1 m condition (middle panel), fatigue yielded (1) a similar increased variance of the CoP displacements to those observed in the No vision and Vision 4 m conditions along the medio-lateral axis (P < 0.01) and (2) a weaker destabilising effect relative to the No vision and Vision 4 m conditions along the anteroposterior axis (P < 0.05).

------------------------------------

Please insert Figure 1 about here

------------------------------------

## 4. Discussion

Given the contradictory findings recently published in the literature [6,7], the present experiment was designed to clarify whether or not visual sensory inputs may compensate for the destabilising effects of calf muscles fatigue during bipedal quiet standing. More precisely, we proposed here to check whether these discrepancies might be attributable to differences between the eye–visual target distances used. To this aim, CoP displacements were measured in 12 healthy subjects, using a force-platform, before and after the performance of a fatiguing calf muscle exercise under three different visual conditions: no vision, and vision of a black cross placed at a distance of either 1 m or 4 m ahead.

In the absence of vision (Fig. 1, left panel), results showed a decreased postural control following calf muscles fatigue, corroborating previous reports [6–10]. If one considers postural control as a perceptual-motor process (e.g., [11]), these features could stem from an alteration of the functionality of the sensory proprioceptive (e.g., [12]) and/or motor systems (e.g., [13]) caused by the fatiguing exercise at the ankle joint. This suggestion is supported by larger destabilising effects observed along the anteroposterior than medio-lateral axis, when considering what the fatiguing exercise involved in terms of joints and tendons receptors stimulation and muscles recruitment (i.e., ankle plantar-flexors muscles).

When vision is available, results suggested that the magnitude of the effect of calf muscles fatigue is dependent on the eye–visual target distance. This observation not only confirms our hypothesis, but also resolves previous divergences regarding the interaction between vision and localised fatigue of calf muscles on postural control during quiet standing. On the one hand, a distant visual target did not allow individuals to limit the destabilising effect induced by a localised fatigue of calf muscles (Vision 4 mcondition, Fig. 1, right panel), in accordance with previous results of Corbeil et al. [6].On the other hand, as recently reported byLedin et al. [7], a nearby visual target was shown to reduce postural sway following calf muscles fatigue with a stabilising effect mostly occurring in the sagittal plane (Vision 1 m condition, Fig. 1, middle panel). As stated in the introduction, these differences could stem fromthe greater resolution in the detection of head motion in the Vision 1 m than Vision 4 m condition, through the geometric consequence of greater retinal shifts of the visual image with a decreased eye–visual target distance. With regard to the hypothesis of an alteration of ankle proprioceptive information induced by muscular fatigue [12], this interpretation is reminiscent with previous results [3–5]. When the reliability of ankle proprioceptive information was reduced by standing on a compliant foam support, visual stabilisation was evidenced to be more efficient with close than distant visual targets. Other authors also showed that individuals were able to take advantage of a nearby visual target to reduce the destabilising effect of a relative muscles weakness of postural muscles – experimentally induced by loading healthy subjects with extra weight additional body weight – observed without vision [7,14]. These findings are in accordance with the results observed in the Vision 1 m condition, assuming the decreased force generating capacity at the ankle joint induced by the fatiguing exercise (e.g., [13]). Finally, it is important to mention that the stabilising effect of vision under conditions of impaired ankle sensory-motor function is not limited to the eye–visual target distance. Rather, other physical and physiological visual parameters, including visual acuity [3,15], contrast sensitivity [15], optical blur [3,16,17], central and peripheral visual fields [3,18], motion parallax [19,20], static and dynamic visual motion cues [21] were shown to affect postural control when standing on a compliant foam support. In the present experiment, these parameters have been controlled to ensure the results observed through this study to only be due to the manipulated variable, i.e., the eye–visual target distance.

In conclusion, results of the present experiment confirmed a decreased postural control consecutive to calf muscles fatigue during quiet bipedal standing in the absence of vision. More interestingly, results also evidenced that the ability to use visual information to compensate for this destabilising effect is dependent on the eye–visual target distance. In the context of the multisensory control of balance, this suggests that the efficiency of the sensory reweighting of visual sensory cues as the neuro-muscular constraints acting on the subject change (e.g., [22,23]) is critically linked with the quality of the information the visual system obtains. More largely, the present results also corroborate previous findings about sensory-motor adaptation processes, i.e., any deficit in one sensory modality is often compensated for by the enhancement of the sensory weights of all the other intact sensory modalities (i.e., not only vision), but that this obviously depends on the relevance of the other sensory cues available in a given environmental context.

**Acknowledgements**

This paper was written while the two first authors were A.T.E.R. at Université de Savoie, France. The authors would like to thank subject volunteers and the anonymous reviewers for helpful comments and suggestions. Special thanks also are extended to Christelle B. and Lucas B. for various contributions.

**Figure captions**

**Figure 1.** Mean and standard error of the mean of the variances of the centre of foot pressure displacements obtained in the two conditions of No fatigue and Fatigue, the three visual conditions of No vision, Vision 1 m and Vision 4 m, along the medio-lateral and antero-posterior axes. The two axes are presented with different symbols: medio-lateral (black diamond) and antero-posterior (white circle). Left, middle and right panels represent No vision, Vision 1 m and Vision 4 m, respectively.

**Figure 1**

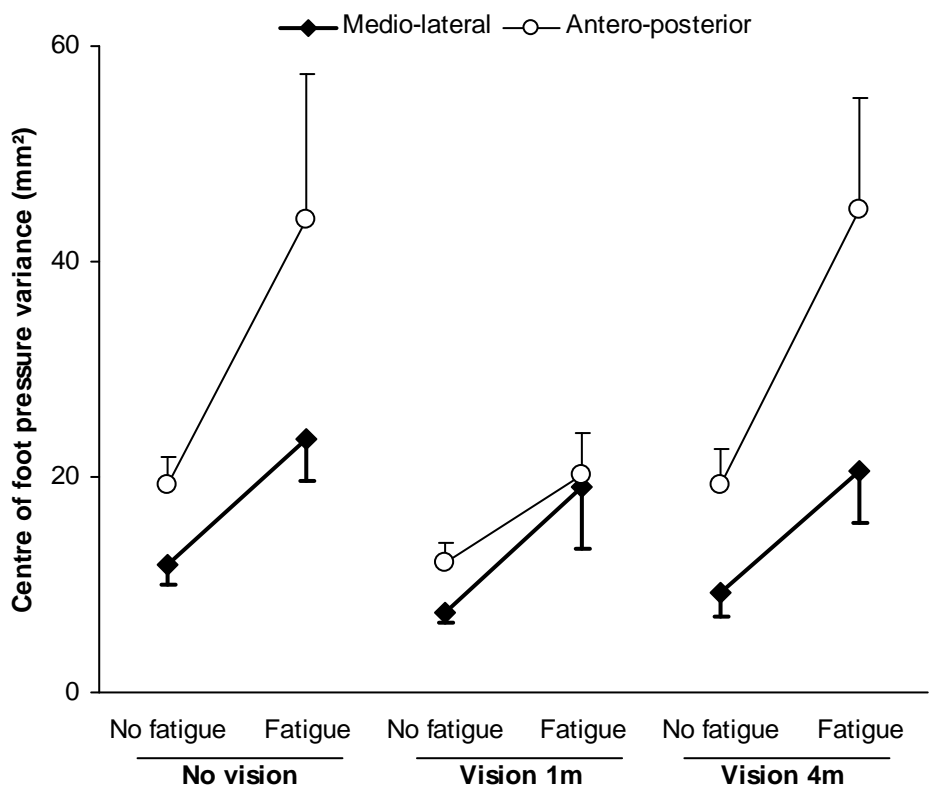